\documentclass[journal]{IEEEtran}
\usepackage{amsmath}
\usepackage{amsfonts}

\newtheorem{theorem}{Theorem}

\newtheorem{definition}[theorem]{Definition}

\newtheorem{remark}[theorem]{Remark}

\input{tcilatex}
\begin{document}

\title{\textbf{The finite harmonic oscillator and its associated sequences%
\thanks{$\overline{\text{Date: January 1, 2007.}}$}\thanks{\copyright \ %
Copyright by S. Gurevich, R. Hadani and N. Sochen, January 1, 2007. All
rights reserved.}}}
\author{\textbf{Shamgar Gurevich, Ronny Hadani, and Nir Sochen}}
\specialpapernotice{To appear in Proceedings of the National Academy of Sciences of the United
States of America}
\maketitle

\begin{abstract}
A novel system of functions (signals) on the finite line, called the
oscillator system, is described and studied. Applications of this system for
discrete radar and digital communication theory are explained.\medskip

\end{abstract}

\section{Introduction}

One-dimensional \textit{analog signals} are complex valued functions on the
real line $%
\mathbb{R}
$. In the same spirit, one-dimensional \textit{digital signals,} also called 
\textit{sequences, }might be considered as complex valued functions on the
finite line $\mathbb{F}_{p}$, i.e., the finite field with $p$ elements. In
both situations the parameter of the line is denoted by $t$ and is referred
to as \textit{time}$.$\textit{\ }In this work, we will consider digital
signals only, which will be simply referred to as signals. The space of
signals $\mathcal{H=}%
\mathbb{C}
(\mathbb{F}_{p})$ is a Hilbert space with the Hermitian product given by 
\begin{equation*}
\left \langle \phi ,\varphi \right \rangle =\tsum \limits_{t\in \mathbb{F}%
p}\phi (t)\overline{\varphi }(t).
\end{equation*}

A central problem is to construct interesting and useful systems of signals.
Given a\ system $\mathfrak{S}$, there are various desired properties which
appear in the engineering wish list. For example, in various situations \cite%
{1, 2} one requires that the signals will be weakly correlated, i.e., that
for every $\phi \neq \varphi \in \mathfrak{S}$ 
\begin{equation*}
\left \vert \left \langle \phi ,\varphi \right \rangle \right \vert \ll 1.
\end{equation*}%
This property is trivially satisfied if $\mathfrak{S}$ is an orthonormal
basis. Such a system cannot consist of more than $\dim (\mathcal{H)}$
signals, however, for certain applications, e.g., CDMA (Code Division
Multiple Access) \cite{3} a larger number of signals is desired, in that
case the orthogonality condition is relaxed.

During the transmission process, a signal $\varphi $ might be distorted in
various ways. Two basic types of distortions are \emph{time shift }$\varphi
(t)\mapsto \mathsf{L}_{\tau }\varphi (t)=\varphi (t+\tau )$ and \emph{phase
shift }$\varphi (t)\mapsto \mathsf{M}_{w}\varphi (t)=e^{\frac{2\pi i}{p}%
wt}\varphi (t)$, where $\tau ,w\in \mathbb{F}_{p}$. The first type appears
in asynchronous communication and the second type is a Doppler effect due to
relative velocity between the transmitting and receiving antennas. In
conclusion, a general distortion is of the type $\varphi \mapsto \mathsf{M}%
_{w}\mathsf{L}_{\tau }\varphi ,$ suggesting that for every $\varphi \neq
\phi \in \mathfrak{S}$ it is natural to require \cite{2} the following
stronger condition 
\begin{equation*}
\left \vert \left \langle \phi ,\mathsf{M}_{w}\mathsf{L}_{\tau }\varphi
\right \rangle \right \vert \ll 1.
\end{equation*}%
Due to technical restrictions in the transmission process, signals are
sometimes required to admit low peak-to-average power ratio \cite{4}, i.e.,
that for every $\varphi \in \mathfrak{S}$ with $\left \Vert \varphi
\right
\Vert _{2}=1$%
\begin{equation*}
\max \left \{ \left \vert \varphi (t)\right \vert :t\in \mathbb{F}_{p}\right
\} \ll 1.
\end{equation*}%
Finally, several schemes for digital communication require that the above
properties will continue to hold also if we replace signals from $\mathfrak{S%
}$ by their Fourier transform$.$

In this paper we construct a novel system of (unit) signals $\mathfrak{S}%
_{O} $, consisting of $\ $order of $p^{3}$ signals, where $p$ is an odd
prime, called the \textit{oscillator system}. These signals constitute, in
an appropriate formal sense, a finite analogue for the eigenfunctions of the
harmonic oscillator in the real setting and, in accordance, they share many
of the nice properties of the latter class. In particular, the system $%
\mathfrak{S}_{O}$ satisfies the following properties

\begin{enumerate}
\item \emph{Autocorrelation (ambiguity function). }For every $\varphi \in 
\mathfrak{S}_{O}$ we have 
\begin{equation}
\left \vert \left \langle \varphi ,\mathsf{M}_{w}\mathsf{L}_{\tau }\varphi
\right \rangle \right \vert =\left \{ 
\begin{array}{c}
1\text{ \  \  \  \  \  \  \ if \ }\left( \tau ,w\right) =0, \\ 
\leq \frac{2}{\sqrt{p}}\text{ \ if }\left( \tau ,w\right) \neq 0.\text{\ }%
\end{array}%
\right.  \label{cross_eq}
\end{equation}

\item \emph{Crosscorrelation (cross-ambiguity function). }For every $\phi
\neq \varphi \in \mathfrak{S}_{O}$ we have 
\begin{equation}
\left \vert \left \langle \phi ,\mathsf{M}_{w}\mathsf{L}_{\tau }\varphi
\right \rangle \right \vert \leq \frac{4}{\sqrt{p}},\   \label{auto_eq}
\end{equation}%
for every $\tau ,w\in \mathbb{F}_{p}$.

\item \emph{Supremum. }For every signal $\varphi \in \mathfrak{S}_{O}$ we
have%
\begin{equation*}
\max \left \{ \left \vert \varphi (t)\right \vert :t\in \mathbb{F}_{p}\right
\} \leq \frac{2}{\sqrt{p}}.
\end{equation*}

\item \emph{Fourier invariance. }For every signal $\varphi \in \mathfrak{S}%
_{O}$ its Fourier transform $\widehat{\varphi }$ is (up to multiplication by
a unitary scalar) also in $\mathfrak{S}_{O}.\ $
\end{enumerate}

\begin{remark}
\bigskip Explicit algorithm that generate the (split) oscillator system is
given in the supporting text.
\end{remark}

The oscillator system can be extended to a much larger system $\mathfrak{S}%
_{E}$, consisting of order of $p^{5}$ signals if one is willing to
compromise Properties 1 and 2 for a weaker condition. The extended system
consists of all signals of the form $\mathsf{M}_{w}\mathsf{L}_{\tau }\varphi 
$ for $\tau ,w\in \mathbb{F}_{p}$ and $\varphi \in \mathfrak{S}_{O}$. It is
not hard to show that $\# \left( \mathfrak{S}_{E}\right) =$ $p^{2}\cdot \#
\left( \mathfrak{S}_{O}\right) \approx p^{5}$. As a consequence of (\ref%
{cross_eq}) and (\ref{auto_eq}) for every $\varphi \neq \phi \in \mathfrak{S}%
_{E}$ we have 
\begin{equation*}
\left \vert \left \langle \varphi ,\phi \right \rangle \right \vert \leq 
\frac{4}{\sqrt{p}}.
\end{equation*}

The characterization and construction of the oscillator system is
representation theoretic and we devote the rest of the paper to an intuitive
explanation of the main underlying ideas. As a suggestive model example we
explain first the construction of the well known system of\ chirp
(Heisenberg) signals, deliberately taking a representation theoretic point
of view (see \cite{5, 2} for a more comprehensive treatment).

\section{Model example (Heisenberg system)}

Let us denote by $\psi :\mathbb{F}_{p}\rightarrow 
\mathbb{C}
^{\times }$ \ the character $\psi (t)=e^{\frac{2\pi i}{p}t}$. We consider
the pair of orthonormal bases $\Delta =\left \{ \delta _{a}:a\in \mathbb{F}%
_{p}\right \} $ and $\Delta ^{\vee }=\left \{ \psi _{a}:a\in \mathbb{F}%
_{p}\right \} $, where $\psi _{a}(t)=\frac{1}{\sqrt{p}}\psi (at),$ and $%
\delta _{a}$ is the Kronecker delta function, $\delta _{a}(t)=1$ if $t=a$
and $\delta _{a}(t)=0$ if $t\neq a.$

\subsection{Characterization of the bases $\Delta $ and $\Delta ^{\vee }$}

Let $\mathsf{L}:\mathcal{H\rightarrow H}$ be the time shift operator $%
\mathsf{L}\varphi (t)=\varphi (t+1)$. This operator is unitary and it
induces a homomorphism of groups $\mathsf{L}:\mathbb{F}_{p}\rightarrow U(%
\mathcal{H)}$ given by $\mathsf{L}_{\tau }\varphi (t)=\varphi (t+\tau )$ for
any $\tau \in \mathbb{F}_{p}$.

Elements of the basis $\Delta ^{\vee }$ are character vectors with respect
to the action $\mathsf{L}$, i.e., $\mathsf{L}_{\tau }\psi _{a}=\psi (a\tau
)\psi _{a}$ for any $\tau \in \mathbb{F}_{p}$. In the same fashion, the
basis $\Delta $ consists of character vectors with respect to the
homomorphism $\mathsf{M}:\mathbb{F}_{p}\rightarrow U(\mathcal{H)}$ given by
the phase shift operators $\mathsf{M}_{w}\varphi (t)=\psi (wt)\varphi (t)$.

\subsection{The Heisenberg representation}

The homomorphisms $\mathsf{L}$ and $\mathsf{M}$ can be combined into a
single map $\widetilde{\pi }:\mathbb{F}_{p}\times \mathbb{F}_{p}\rightarrow
U(\mathcal{H)}$ which sends a pair $(\tau ,w)$ to the unitary operator $%
\widetilde{\pi }(\tau ,\omega )=\psi \left( -\tfrac{1}{2}\tau w\right) 
\mathsf{M}_{w}\circ \mathsf{L}_{\tau }$. The plane $\mathbb{F}_{p}\times 
\mathbb{F}_{p}$ is called the \textit{time-frequency plane} and will be
denoted by $V$. The map $\widetilde{\pi }$ is not an homomorphism since, in
general, the operators $L_{\tau }$ and $M_{w}$ do not commute. This
deficiency can be corrected if we consider the group $H=V\times \mathbb{F}%
_{p}$ with multiplication given by 
\begin{equation*}
(\tau ,w,z)\cdot (\tau ^{\prime },w^{\prime },z^{\prime })=(\tau +\tau
^{\prime },w+w^{\prime },z+z^{\prime }+\tfrac{1}{2}(\tau w^{\prime }-\tau
^{\prime }w)).
\end{equation*}%
The map $\widetilde{\pi }$ extends to a homomorphism $\pi :H\rightarrow U(%
\mathcal{H)}$ given by 
\begin{equation*}
\pi (\tau ,w,z)=\psi \left( -\tfrac{1}{2}\tau w+z\right) \mathsf{M}_{w}\circ 
\mathsf{L}_{\tau }.
\end{equation*}%
The group $H$ is called the \textit{Heisenberg }group and the homomorphism $%
\pi $ is called the \textit{Heisenberg representation}$.$

\subsection{Maximal commutative subgroups}

The Heisenberg group is no longer commutative, however, it contains various
commutative subgroups which can be easily described. To every line $L\subset
V,$ that pass through the origin, one can associate a maximal commutative
subgroup $A_{L}=\left \{ (l,0)\in V\times \mathbb{F}_{p}:l\in L\right \} $.
It will be convenient to identify the subgroup $A_{L}$ with the line $L$.

\subsection{Bases associated with lines}

Restricting the Heisenberg representation $\pi $ to a subgroup $L$ yields a
decomposition of the Hilbert space $\mathcal{H}$ into a direct sum of
one-dimensional subspaces $\mathcal{H=}\tbigoplus \limits_{\chi }\mathcal{H}%
_{\chi },$ where $\chi $ runs in the set $L^{\vee }$ of (complex valued)
characters of the group $L$. The subspace $\mathcal{H}_{\chi }$ consists of
vectors $\varphi \in \mathcal{H}$ such that $\pi (l)\varphi =\chi (l)\varphi 
$. In other words, the space $\mathcal{H}_{\chi }$ consists of common
eigenvectors with respect to the commutative system of unitary operators $%
\left \{ \pi (l)\right \} _{l\in L}$ such that the operator $\pi \left(
l\right) $ has eigenvalue $\chi \left( l\right) $.

Choosing a unit vector $\varphi _{\chi }\in \mathcal{H}_{\chi \text{ }}$for
every $\chi \in L^{\vee }$ \ we obtain an orthonormal basis $\mathcal{B}%
_{L}=\left \{ \varphi _{\chi }:\chi \in L^{\vee }\right \} $. In particular, 
$\Delta ^{\vee }$ and $\Delta $ are recovered as the bases associated with
the lines $T=\left \{ (\tau ,0):\tau \in \mathbb{F}_{p}\right \} $ and $%
W=\left \{ (0,w):w\in \mathbb{F}_{p}\right \} $ respectively. For a general $%
L$ the signals in $\mathcal{B}_{L}$ are certain kind of chirps. Concluding,
we associated with every line $L\subset V$ \ an orthonormal basis $\mathcal{B%
}_{L},$ and overall we constructed a system of signals consisting of a union
of orthonormal bases 
\begin{equation*}
\mathfrak{S}_{H}\mathfrak{=}\left \{ \varphi \in \mathcal{B}_{L}:L\subset
V\right \} .
\end{equation*}%
For obvious reasons, the system $\mathfrak{S}_{H}$ will be called the 
\textit{Heisenberg }system\textit{. }

\subsection{Properties of the Heisenberg system}

\smallskip It will be convenient to introduce the following general notion.
Given two signals $\phi ,\varphi \in \mathcal{H}$, their matrix coefficient
is the function $m_{\phi ,\varphi }:H\rightarrow 
\mathbb{C}
$ \ given by $m_{\phi ,\varphi }(h)=\left \langle \phi ,\pi (h)\varphi
\right \rangle $. In coordinates, if we write $h=\left( \tau ,w,z\right) $
then $m_{\phi ,\varphi }(h)=\psi \left( -\tfrac{1}{2}\tau w+z\right)
\left
\langle \phi ,\mathsf{M}_{w}\circ \mathsf{L}_{\tau }\varphi
\right
\rangle $. When $\phi =\varphi $ the function $m_{\varphi ,\varphi }$
is called the \textit{ambiguity} function of the vector $\varphi $ and is
denoted by $A_{\varphi }=m_{\varphi ,\varphi }$.\smallskip \ 

The system $\mathfrak{S}_{H}$ consists of $p+1$ orthonormal bases\footnote{%
Note that $p+1$ is the number of lines in $V$.}, altogether $p\left(
p+1\right) $ signals and it satisfies the following properties \cite{5, 2}

\begin{enumerate}
\item \emph{Autocorrelation}\textbf{. }For every signal $\varphi \in 
\mathcal{B}_{L}$ the function $|A_{\varphi }|$ is the characteristic
function of the line $L$, i.e., 
\begin{equation*}
\left \vert A_{\varphi }\left( v\right) \right \vert =\left \{ 
\begin{array}{c}
0,\text{ \ }v\notin L, \\ 
1,\text{\  \ }v\in L.%
\end{array}%
\right.
\end{equation*}

\item \emph{Crosscorrelation}.\textbf{\ }For every $\phi \in \mathcal{B}_{L}$
and $\varphi \in \mathcal{B}_{M}$ where $L\neq M$ we have%
\begin{equation*}
\left \vert m_{\varphi ,\phi }\left( v\right) \right \vert \leq \frac{1}{%
\sqrt{p}},
\end{equation*}%
for every $v\in V$. If $L=M$ then $m_{\varphi ,\phi }$ is the characteristic
function of some translation of the line $L$.

\item \emph{Supremum}\textbf{. }A signal $\varphi \in \mathfrak{S}_{H}$ is a
unimodular function, i.e., $\left \vert \varphi (t)\right \vert =\frac{1}{%
\sqrt{p}}$ for every $t\in \mathbb{F}_{p}$, in particular we have 
\begin{equation*}
\max \left \{ \left \vert \varphi (t)\right \vert :t\in \mathbb{F}_{p}\right
\} =\frac{1}{\sqrt{p}}\ll 1\text{.}
\end{equation*}
\end{enumerate}

\begin{remark}
Note the main differences between the Heisenberg and the oscillator systems.
The oscillator system consists of order of $p^{3}$ signals, while the
Heisenberg system consists of \ order of $p^{2}$ signals. Signals in the
oscillator system admits an ambiguity function concentrated at $0\in V$
(thumbtack pattern) while signals in the Heisenberg system admits ambiguity
function concentrated on a line.
\end{remark}

\section{The oscillator system}

Reflecting back on the Heisenberg system we see that each vector $\varphi
\in \mathfrak{S}_{H}$ is characterized in terms of action of the additive
group $G_{a}=\mathbb{F}_{p}$. Roughly, in comparison, each vector in the
oscillator system is characterized in terms of action of the multiplicative
group $G_{m}=\mathbb{F}_{p}^{\times }$. Our next goal is to explain the last
assertion. We begin by giving a model example.

Given a multiplicative character\footnote{%
A multiplicative character is a function $\chi :G_{m}\rightarrow 
\mathbb{C}
^{\times }$ which satisfies $\chi (xy)=\chi (x)\chi (y)$ for every $x,y\in
G_{m}.$} $\chi :G_{m}\rightarrow 
\mathbb{C}
^{\times }$, we define a vector $\underline{\chi }\in \mathcal{H}$ by 
\begin{equation*}
\underline{\chi }(t)=\left \{ 
\begin{array}{c}
\frac{1}{\sqrt{p-1}}\chi (t),\text{ \  \  \ }t\neq 0, \\ 
0,\text{ \  \  \  \  \  \  \  \  \  \  \  \  \  \  \ }t=0.%
\end{array}%
\right.
\end{equation*}%
We consider the system $\mathcal{B}_{std}=\left \{ \underline{\chi }:\chi
\in G_{m}^{\vee },\text{ }\chi \neq 1\right \} $, where $G_{m}^{\vee }$ is
the dual group of characters.

\subsection{Characterizing the system $\mathcal{B}_{std}$}

For each element $a\in G_{m}$ let $\  \rho _{a}:\mathcal{H\rightarrow H}$ \
be the unitary operator acting by scaling $\rho _{a}\varphi (t)=\varphi (at)$%
. This collection of operators form a homomorphism $\rho :G_{m}\rightarrow U(%
\mathcal{H)}$.

Elements of $\mathcal{B}_{std}$ are character vectors with respect to $\rho $%
, i.e., the vector$\underline{\text{ }\chi }$ satisfies $\rho _{a}\left( 
\underline{\chi }\right) =\chi (a)\underline{\chi }$ for every $a\in G_{m}$.
In more conceptual terms, the action $\rho $ yields a decomposition of the
Hilbert space $\mathcal{H}$ into character spaces $\mathcal{H=}\tbigoplus 
\mathcal{H}_{\chi }$, where $\chi $ runs in the group $G_{m}^{\vee }$. The
system $\mathcal{B}_{std}$ consists of a representative unit vector for each
space $\mathcal{H}_{\chi }$, $\chi \neq 1$.

\subsection{The Weil representation}

We would like to generalize the system $\mathcal{B}_{std}$ in a similar
fashion like we generalized the bases $\Delta $ and $\Delta ^{\vee }$ in the
Heisenberg setting. In order to do this we need to introduce several
auxiliary operators.

Let $\rho _{a}:\mathcal{H\rightarrow H}$, $a\in \mathbb{F}_{p}^{\times },$
be the operators acting by $\rho _{a}\varphi (t)=\sigma (a)\varphi (a^{-1}t)$
(scaling), where $\sigma $ is the unique quadratic character of $\mathbb{F}%
_{p}^{\times }$, let $\rho _{T}:\mathcal{H\rightarrow H}$ be the operator
acting by $\rho _{T}\varphi (t)=\psi (t^{2})\varphi (t)$ (quadratic
modulation), and finally let $\rho _{S\text{ }}:\mathcal{H\rightarrow H}$ be
the operator of Fourier transform 
\begin{equation*}
\rho _{S}\varphi (t)=\frac{\nu }{\sqrt{p}}\tsum_{s\in \mathbb{F}_{p}}\psi
(ts)\varphi (s),
\end{equation*}%
where $\nu $ is a normalization constant \cite{6}. The operators $\rho
_{a},\rho _{T}$ and $\rho _{S}$ are unitary. Let us consider the subgroup of
unitary operators generated by $\rho _{a},\rho _{S}$ and $\rho _{T}$. This
group turns out to be isomorphic to the finite group $Sp=SL_{2}(\mathbb{F}%
_{p})$, therefore we obtained a homomorphism $\rho :Sp\rightarrow U(\mathcal{%
H)}$. The representation $\rho $ is called the \textit{Weil representation} 
\cite{7} and it will play a prominent role in this paper.

\subsection{Systems associated with maximal (split) tori}

The group $Sp$ consists of various types of commutative subgroups. We will
be interested in maximal \emph{diagonalizable} commutative subgroups. A
subgroup of this type is called maximal \textit{split} \textit{torus. }The
standard example is the subgroup consisting of all diagonal matrices 
\begin{equation*}
A=\left \{ 
\begin{pmatrix}
a & 0 \\ 
0 & a^{-1}%
\end{pmatrix}%
:a\in G_{m}\right \} ,
\end{equation*}%
which is called the \textit{standard torus}. The restriction of the Weil
representation to a split torus $T\subset Sp$ yields a decomposition of the
Hilbert space $\mathcal{H}$ into a direct sum of character spaces $\mathcal{%
H=}\tbigoplus \mathcal{H}_{\chi }$, where $\chi $ runs in the set of
characters $T^{\vee }$. Choosing a unit vector $\varphi _{\chi }\in \mathcal{%
H}_{\chi \text{ }}$ for every $\chi $ we obtain a collection of orthonormal
vectors $\mathcal{B}_{T}=\left \{ \varphi _{\chi }:\chi \in T^{\vee },\text{ 
}\chi \neq \sigma \right \} $. Overall, we constructed a system 
\begin{equation*}
\mathfrak{S}_{O}^{s}\mathfrak{=}\left \{ \varphi \in \mathcal{B}%
_{T}:T\subset Sp\text{ split}\right \} ,
\end{equation*}%
which will be referred to as the \textit{split oscillator system. We note
that }our initial system $\mathcal{B}_{std}$ \ is recovered as $\mathcal{B}%
_{std}=\mathcal{B}_{A}$.

\subsection{Systems associated with maximal (non-split) tori}

From the point of view of this paper, the most interesting maximal
commutative subgroups in $Sp$ are those which are diagonalizable over an
extension field rather than over the base field $\mathbb{F}_{p}$. A subgroup
of this type is called maximal \textit{non-split torus. }It might be
suggestive to first explain the analogue notion in the more familiar setting
of the field $%
\mathbb{R}
$. \ Here, the standard example of a maximal non-split torus is the circle
group $SO(2)\subset SL_{2}(%
\mathbb{R}
)$. Indeed, it is a maximal commutative subgroup which becomes
diagonalizable when considered over the extension field $%
\mathbb{C}
$ of complex numbers.

The above analogy suggests a way to construct examples of maximal non-split
tori in the finite field setting as well. Let us assume for simplicity that $%
-1$ does not admit a square root in $\mathbb{F}_{p}$. The group $Sp$ acts
naturally on the plane $V=\mathbb{F}_{p}\times \mathbb{F}_{p}$. Consider the
symmetric bilinear form $B$ on $V$ given by 
\begin{equation*}
B((t,w),(t^{\prime },w^{\prime }))=tt^{\prime }+ww^{\prime }.
\end{equation*}

An example of maximal non-split torus is the subgroup $T_{ns}\subset Sp$
consisting of all elements $g\in Sp$ preserving the form $B$, i.e., $g\in
T_{ns}$ if and only if $B(gu,gv)=B(u,v)$ for every $u,v\in V$. In the same
fashion like in the split case, restricting the Weil representation to a
non-split torus $T$ yields a decomposition into character spaces $\mathcal{H=%
}\tbigoplus \mathcal{H}_{\chi }$. Choosing a unit vector $\varphi _{\chi
}\in \mathcal{H}_{\chi }$ for every $\chi \in T^{\vee }$ we obtain an
orthonormal basis $\mathcal{B}_{T}$. Overall, we constructed a system of
signals 
\begin{equation*}
\mathfrak{S}_{O}^{ns}\mathfrak{=}\left \{ \varphi \in \mathcal{B}%
_{T}:T\subset Sp\text{ non-split}\right \} .
\end{equation*}%
The system $\mathfrak{S}_{O}^{ns}$ will be referred to as the \textit{%
non-split oscillator }system\textit{. }The construction of the system $%
\mathfrak{S}_{O}=$ $\mathfrak{S}_{O}^{s}\cup \mathfrak{S}_{O}^{ns}$ together
with the formulation of some of its properties are the main contribution of
this paper.

\subsection{Behavior under Fourier transform}

The oscillator system is closed under the operation of Fourier transform,
i.e., for every $\varphi \in \mathfrak{S}_{O}$ we have $\widehat{\varphi }%
\in \mathfrak{S}_{O}.$ The Fourier transform on the space $%
\mathbb{C}
\left( \mathbb{F}_{p}\right) $ appears as a specific operator $\rho \left( 
\mathrm{w}\right) $ in the Weil representation, where 
\begin{equation*}
\mathrm{w}=%
\begin{pmatrix}
0 & 1 \\ 
-1 & 0%
\end{pmatrix}%
\in Sp.
\end{equation*}%
Given a signal $\varphi \in \mathcal{B}_{T}\subset \mathfrak{S}_{O}$, its
Fourier transform $\widehat{\varphi }=\rho \left( \mathrm{w}\right) \varphi $
is, up to a unitary scalar, a signal in $\mathcal{B}_{T^{\prime }}$ where $%
T^{\prime }=\mathrm{w}T\mathrm{w}^{-1}$ . In fact, $\mathfrak{S}_{O}$ is
closed under all the operators in the Weil representation! Indeed, given an
element $g\in Sp$ and a signal $\varphi \in \mathcal{B}_{T}$ we have, up to
a unitary scalar, that $\rho \left( g\right) \varphi $ $\in \mathcal{B}%
_{T^{\prime }}$, where $T^{\prime }=gTg^{-1}$.

In addition, the Weyl element $\mathrm{w}$ is an element in some maximal
torus $T_{\mathrm{w}}$ (the split type of $T_{\mathrm{w}}$ depends on the
characteristic $p$ of the field) and as a result signals $\varphi \in 
\mathcal{B}_{T_{\mathrm{w}}}$ are, in particular, eigenvectors of the
Fourier transform. As a consequences a signal $\varphi \in \mathcal{B}_{T_{%
\mathrm{w}}}$ and its Fourier transform $\widehat{\varphi }$ differ by a
unitary constant, therefore are practically the "same" for all essential
matters.

These properties might be relevant for applications to OFDM (Orthogonal
Frequency Division Multiplexing) \cite{8} where one requires good properties
both from the signal and its Fourier transform.

\subsection{Relation to the harmonic oscillator}

Here we give the explanation why functions in the non-split oscillator
system $\mathfrak{S}_{O}^{ns}$ constitute a finite analogue of the
eigenfunctions of the harmonic oscillator in the real setting. The Weil
representation establishes the dictionary between these two, seemingly,
unrelated objects. The argument works as follows.

The one-dimensional harmonic oscillator is given by the differential
operator $D=\partial ^{2}-t^{2}$. The operator $D$ can be exponentiated to
give a unitary representation of the circle group $\rho :SO\left( 2,%
\mathbb{R}
\right) \longrightarrow U\left( L^{2}(%
\mathbb{R}
\right) )$ where $\rho (\theta )=e^{i\theta D}$. Eigenfunctions of $D$ are
naturally identified with character vectors with respect to $\rho $. The
crucial point is that $\rho $ is the restriction of the Weil representation
of $SL_{2}\left( 
\mathbb{R}
\right) $ to the maximal non-split torus $SO\left( 2,%
\mathbb{R}
\right) \subset SL_{2}\left( 
\mathbb{R}
\right) $.

Summarizing, the eigenfunctions of the harmonic oscillator and functions in $%
\mathfrak{S}_{O}^{ns}$ are governed by the same mechanism, namely both are
character vectors with respect to the restriction of the Weil representation
to a maximal non-split torus in $SL_{2}$. The only difference appears to be
the field of definition, which for the harmonic oscillator is the reals and
for the oscillator functions is the finite field.

\section{Applications}

Two applications of the oscillator system will be described. The first
application is to the theory of discrete radar. The second application is to
CDMA systems. We will give a brief explanation of these problems, while
emphasizing the relation to the Heisenberg representation.

\subsection{Discrete Radar}

The theory of discrete radar is closely related \cite{2} to the finite
Heisenberg group $H.$ A radar sends a signal $\varphi (t)$ and obtains an
echo $e(t)$. The goal \cite{9} is to reconstruct, in maximal accuracy, the
target range and velocity. The signal $\varphi (t)$ and the echo $e(t)$ are,
principally, related by the transformation%
\begin{equation*}
e(t)=e^{2\pi iwt}\varphi (t+\tau )=\mathsf{M}_{w}\mathsf{L}_{\tau }\varphi
(t),
\end{equation*}%
where the time shift $\tau $ encodes the distance of the target from the
radar and the phase shift encodes the velocity of the target. Equivalently
saying, the transmitted signal $\varphi $ and the received echo $e$ are
related by an action of an element $h_{0}\in H$, i.e., $e=\pi (h_{0})\varphi
.$ The problem of discrete radar can be described as follows. Given a signal 
$\varphi $ and an echo $e=\pi (h_{0})\varphi $ extract the value of $h_{0}$%
\textbf{.} \ 

It is easy to show that $\left \vert m_{\varphi ,e}\left( h\right)
\right
\vert =\left \vert A_{\varphi }\left( h\cdot h_{0}\right)
\right
\vert $ and it obtains its maximum at $h_{0}^{-1}$. This suggests
that a desired signal $\varphi $ for discrete radar should admit an
ambiguity function $A_{\varphi } $ which is highly concentrated around $0\in
H$, which is a property satisfied by signals in the oscillator system
(Property 2).

\begin{remark}
It should be noted that the system $\mathfrak{S}_{O}$ is \ "large"
consisting of $\ $aproximately $p^{3}$ signals. This property becomes
important in a \textit{jamming }scenario.
\end{remark}

\subsection{Code Division Multiple Access (CDMA)}

We are considering the following setting.

\begin{itemize}
\item There exists a collection of users $i\in I$, each holding a \textit{bit%
} of information $b_{i}\in 
\mathbb{C}
$ \ (usually $b_{i}$ is taken to be an $N$'th root of unity).

\item Each user transmits his bit of information, say, to a central antenna.
In order to do that, \ he multiplies his bit $b_{i}$ by a private signal $%
\varphi _{i}\in \mathcal{H}$ \ and forms a message $u_{i}=b_{i}\varphi _{i}$.

\item The transmission is carried through a single channel (for example in
the case of cellular communication the channel is the atmosphere), therefore
the message received by the antenna is the sum%
\begin{equation*}
u=\tsum \limits_{i}u_{i}.
\end{equation*}
\end{itemize}

The main problem \cite{3} is to extract the individual bits $b_{i}$ from the
message $u$. The bit $b_{i}$ can be estimated by calculating the inner
product%
\begin{equation}
\left \langle \varphi _{i},u\right \rangle =\tsum \limits_{j}\left \langle
\varphi _{i},u_{j}\right \rangle =\tsum \limits_{j}b_{j}\left \langle
\varphi _{i},\varphi _{j}\right \rangle =b_{i}+\tsum \limits_{j\neq
i}b_{j}\left \langle \varphi _{i},\varphi _{j}\right \rangle .  \notag
\end{equation}%
The last expression above should be considered as a sum of the information
bit $b_{i}$ and an additional noise caused by the interference\textit{\ }of
the other messages. This is the standard scenario also called the \textit{%
Synchronous} scenario. In practice, more complicated scenarios appear, e.g., 
\emph{asynchronous scenario }- in which\emph{\ }each message $u_{i}$ is
allowed to acquire an arbitrary time shift $u_{i}(t)\mapsto u_{i}(t+\tau
_{i})$, \emph{phase shift scenario} - in which each message $u_{i}$ is
allowed to acquire an arbitrary phase shift $u_{i}(t)\mapsto e^{\frac{2\pi i%
}{p}w_{i}t}u_{i}(t)$ and probably also a combination of the two where each
message $u_{i}$ is allowed to acquire an arbitrary distortion of the form $%
u_{i}(t)\mapsto e^{\frac{2\pi i}{p}w_{i}t}u_{i}(t+\tau _{i}).$

The previous discussion suggests that what we are seeking for is a large
system $\mathfrak{S}$ of signals which will enable a reliable extraction of
each bit $b_{i}$ for as many users transmitting through the channel
simultaneously.

\begin{definition}[Stability conditions]
\label{stability_def}\smallskip \ Two unit signals $\phi \neq $ $\varphi $
are called \textbf{stably cross-correlated} if $\left \vert m_{\varphi ,\phi
}\left( v\right) \right \vert \ll 1$ for every $v\in V$. A unit signal $%
\varphi $ is called \textbf{stably autocorrelated\ }if $\left \vert
A_{\varphi }\left( v\right) \right \vert \ll 1$, for every $v\neq 0$. A
system $\mathfrak{S}$ of signals is called a \textbf{stable}\ system if
every signal $\varphi \in \mathfrak{S}$ is stably autocorrelated and any two
different signals $\phi ,\varphi \in \mathfrak{S}$ are stably
cross-correlated.
\end{definition}

Formally what we require for CDMA is a stable system $\mathfrak{S}$. Let us
explain why this corresponds to a reasonable solution to our problem. At a
certain time $t$ the antenna receives a message 
\begin{equation*}
u=\tsum \limits_{i\in J}u_{i},
\end{equation*}%
which is transmitted from a subset of users $J\subset I$. Each message $%
u_{i} $, $i\in J,$ $\ $is of the form $u_{i}=b_{i}e^{\frac{2\pi i}{p}%
w_{i}t}\varphi _{i}(t+\tau _{i})=b_{i}\pi (h_{i})\varphi _{i},$ where $%
h_{i}\in H$. \ In order to extract the bit $b_{i}$ we compute the matrix
coefficient%
\begin{equation*}
m_{\varphi _{i},u}=b_{i}R_{h_{i}}A_{\varphi _{i}}+\#(J-\{i\})o(1),
\end{equation*}

where $R_{h_{i}}$ is the operator of right translation $R_{h_{i}}A_{\varphi
_{i}}(h)=A_{\varphi _{i}}(hh_{i}).$

If the cardinality of the set $J$ is not too big then by evaluating $%
m_{\varphi _{i},u}$ at $h=h_{i}^{-1}$ we can reconstruct the bit $b_{i}$. It
follows from (\ref{cross_eq}) and (\ref{auto_eq}) that the oscillator system 
$\mathfrak{S}_{O\text{ }}$can support order of $p^{3}$ users, enabling
reliable reconstruction when order of $\sqrt{p}$ users are transmitting
simultaneously.

\textbf{Remark about field extensions. }All the results in this paper were
stated for the basic finite field $\mathbb{F}_{p}$ for the reason of making
the terminology more accessible. However, they are valid for any field
extension of the form $\mathbb{F}_{q}$ with $q=p^{n}.$ Complete proofs
appear in \cite{6}.

\textbf{Acknowledgement.} The authors would like to thank J. Bernstein for
his interest and guidance in the mathematical aspects of this work. We are
grateful to S. Golomb and G. Gong for their interest in this project. We
thank B. Sturmfels for encouraging us to proceed in this line of research.
The authors would like to thank V. Anantharam, A. Gr\"{u}nbaum and A. Sahai
for interesting discussions. Finally, the second author is indebted to B.
Porat for so many discussions where each tried to understand the cryptic
terminology of the other.


\begin{thebibliography}{1}
\bibitem[1]{1} Golomb S.W. and Gong G., Signal design for good correlation.
For wireless communication, cryptography, and radar. \textit{Cambridge
University Press, Cambridge (2005).}

\bibitem[2]{2} Howard S. D., Calderbank A. R. and Moran W., The finite
Heisenberg-Weyl groups in radar and communications. \textit{URASIP Journal
on Applied Signal Processing Volume 2006 (2006), Article ID 85685, 12 pages.}

\bibitem[3]{3} Viterbi A.J., CDMA: Principles of Spread Spectrum
Communication. \textit{Addison-Wesley (1995)}.

\bibitem[4]{4} Paterson, K.G. and Tarokh V., On the existence and
construction of good codes with low peak-to-average power ratios. \textit{%
IEEE Trans. Inform. Theory 46 (2000) 1974-1987.}

\bibitem[5]{5} Howe R., Nice error bases, mutually unbiased bases, induced
representations, the Heisenberg group and finite geometries. \textit{Indag.
Math. (N.S.) 16 (2005), no. 3-4, 553--583.}

\bibitem[6]{6} Gurevich S., Hadani R. and Sochen N., The finite harmonic
oscillator and its applications to sequences, communication and radar. 
\textit{IEEE Transactions on Information Theory, in press (2008).}

\bibitem[7]{7} Weil A., Sur certains groupes d'operateurs unitaires. \textit{%
Acta Math. 111} (1964) \textit{143-211.}

\bibitem[8]{8} Chang R.W., Synthesis of Band-Limited Orthogonal Signals for
Multichannel Data Transmission. \textit{Bell System Technical Journal 45
(1966) 1775-1796}.

\bibitem[9]{9} Woodward P.M., Probability and Information theory, with
Applications to Radar. \textit{Pergamon Press, New York} \textit{(1953).}
\end{thebibliography}
\end{document}